      [1999/12/01 v1.4c Il Nuovo Cimento]
\documentclass{cimento}

\usepackage{graphicx}  
\title{Large Scale Structures at High Redshift in the GOODS Field}
\author{M.~Castellano\from{ins:x}\ETC, S.~Salimbeni\from{ins:y}, D.~Trevese\from{ins:x},  L.~Pentericci\from{ins:y}, A.~Grazian\from{ins:y}, A.~Fontana\from{ins:y}, E.~Giallongo\from{ins:y}, P.~Santini\from{ins:y}, S.~Cristiani\from{ins:z}, M.~Nonino\from{ins:z}  \atque E.~Vanzella\from{ins:z} 
}

\instlist{\inst{ins:x} 1--Dipartimento di Fisica, Università di Roma “La Sapienza”, P.le A. Moro 2, 00185 Roma, Italy
\inst{ins:y} INAF - Osservatorio Astronomico di Roma, Via di Frascati 33, 00040 Monte Porzio Catone, Italy 
\inst{ins:z}INAF - Osservatorio Astronomico di Trieste, Via G.B. Tiepolo 11, 34131 Trieste, Italy}

\PACSes{\PACSit{98.65.Cw}{Galaxy clusters}}
\begin{document}

\maketitle

\begin{abstract}

We present a catalogue of overdensities in the GOODS-South field. We find various high density peaks that are embedded in structures diffused on the entire field, up to $z\sim$2.5. The slope of their colour-magnitude relation does not show significative evolution with $z$. We find evidence that galaxies forming these structures are more massive than galaxies located in low density regions. We also analyse the variation of galaxy properties with the associated environmental density and we find that the segregation of red galaxies with density is stronger at low redshift and at high luminosities while it gets much weaker for increasing $z$.

\end{abstract}

\section{Catalogue and Cluster Finding Algorithm}

The GOODS-MUSIC catalogue \cite{grazian} covers 143 $arcmin^{2}$ over the Chandra Deep Field South.
It contains photometry in 14 bands for 14847 extra-galactic objects selected with $Z<26.18$  or with $Ks<23.8$.

About 10\% of the objects have been observed spectroscopically. Where spectroscopy was not available we have used photometric redshifts obtained from our photometric redshift code (e.g. \cite{fontana00}) that employs a standard $\chi^{2}$ minimization over a large set of templates obtained from synthetic spectral models. We present the application to the GOODS-MUSIC catalogue in the interval $0.4<z<2.4$ of a simple  method to evaluate  galaxy volume density through the use of photo-z \cite{trevese07}. Here we outline its relevant steps:

- We divide the survey volume in cells whose extension in different directions depends on the relevant positional accuracy (the angular positions being more accurate than distances).

- For each cell in space we then count neighboring objects at increasing  distance, until a number n of objects is reached. We then assign to the cell a comoving density  $\rho= n/V_{n}$.

- We take into account the brightening of limiting absolute magnitude with increasing redshift for a given apparent magnitude limit, assigning a weight to each detected galaxy at redshift $z$.  

- Galaxy overdensities of any shape (groups, walls, filaments etc.), are defined as  connected 3-dimensional regions with density exceeding a fixed threshold. Cells are grouped together according to a FoF technique.

Throughout the paper we adopt $\Omega_{\Lambda}=0.7, ~ \Omega_M  =0.3, ~  H_0=70~ km ~ s^{-1} ~ Mpc^{-1}$. 

\begin{table}
   \caption{Structures in the GOODS-South Field}
   \label{tab:structures}
   \begin{tabular}{rcl}
Redshift & N Obj. & References \\
     \hline
       0.65 - 0.75 &  123 &  Gilli et al. 2003, Adami et al. 2005, Trevese et al. 2007   \\
       1.0  - 1.1 & 108 &  Adami et al. 2005, Trevese et al. 2007, Di\`{a}z Sa\'{n}chez et al. 2007   \\
       1.6& 45  & Castellano et al. 2007 \\
      2.2 - 2.4   & 113 &   \\
     \hline
   \end{tabular}
 \end{table}

\section{Catalogue of Overdensities and Effects of the Environment at Different Redshifts}

\begin{figure}
\includegraphics[width=6cm]{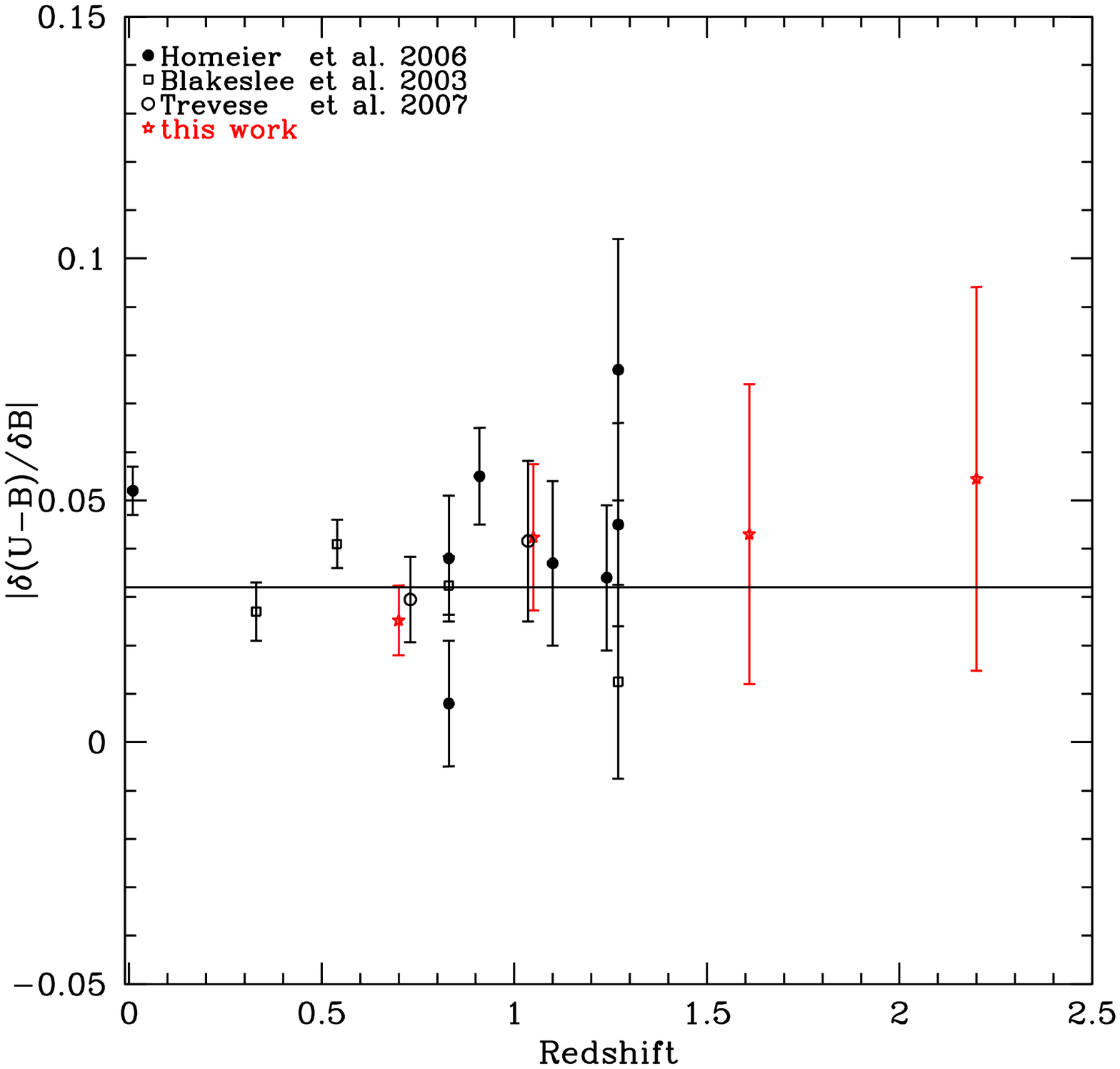}
\includegraphics[width=6cm]{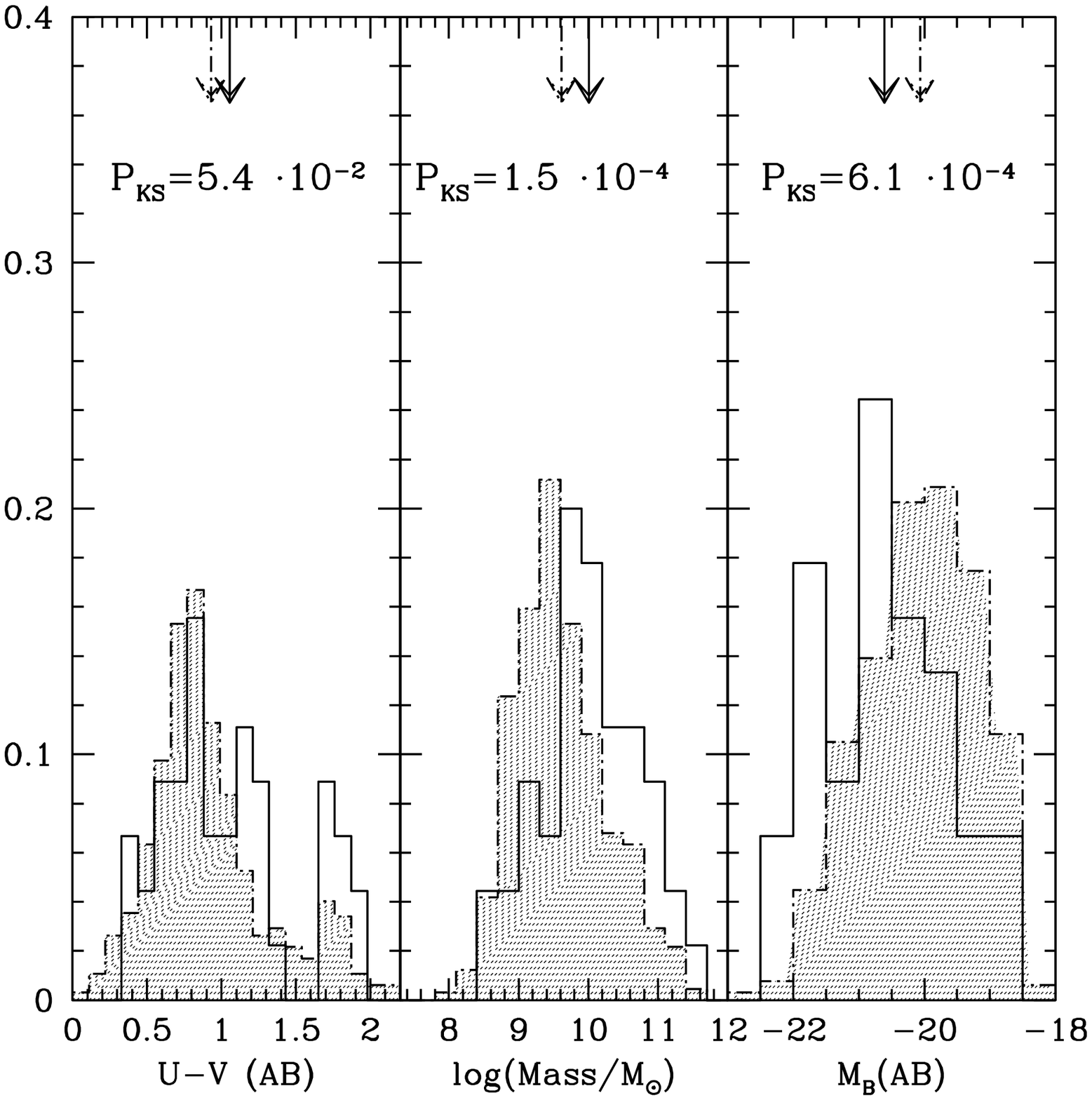}
\caption{Left: The slope of the 'red sequence' for structures in the GOODS field (red points) and for clusters taken from the literature. Right: Fraction of galaxies according to the (U-V) color, total stellar mass (in solar units) and $M_B$ (AB) magnitude: objects are selected in low density regions (shaded histogram) or in the cluster at $z\sim$1.6. In each panel we indicate the Kolmogorv-Smirnov probability of the null hypothesis that the two samples are drawn from the same distribution. The averages of the distributions are indicated with arrows}
\end{figure}
We have analyzed galaxy properties in two complementary ways: through the detection and study of the highest density peaks and analysing their variation with the environmental density. We find overdensities at various redshifts that are spread over the entire field. They  are presented in Table 1 along with references to previous identifications of the structure in the literature  \cite{gilli2003,adami05,diaz07}. Using a threshold of 4$\sigma$ above the average, we can isolate some peaks. We then assign to the peaks all objects, within an Abell radius, with an associated environmental density of at least 2$\sigma$ above the average.

We calculate the (U-B) vs B colour-magnitude diagrams considering all the overdensities at the following redshifts: 0.7, 1.0, 1.6, 2.3. ‘Red sequence’ galaxies are chosen as those fitted by a model with t/$\tau >$ 4, where t is the age of the object and$\ \tau$ its star-formation e-folding time.
The slope of this 'red sequence' does not show any significant variation with redshift, as shown in Fig. 1. This implies that the mass-metallicity relation that produces the red sequence  remains constant with redshift.

The most interesting structure we detected is a diffuse one, distributed over the entire GOODS field around $z$=1.61. Within this extended structure we isolated a compact, higher density peak that we identify as a cluster \cite{me}. We selected a sample of 45 ‘cluster ’ galaxies  ($\rho = \bar{\rho} +2\sigma_{\rho}$ and  $1.45<z<1.75$ , i.e. z=$1.6 \pm2\sigma_z$) and a sample of galaxies located in low density regions, ( $\rho <  \bar{\rho}$  and outside  2 Abell radii RA from the overdensity peak). 
From the distributions of rest frame (U-V) colors, total stellar mass and rest frame $M_B$ magnitudes we can see that the galaxies in the overdensity are significantly brighter and more massive than the  galaxies at low $\rho$(second panel in Fig. 1). Although the overdensity is mainly formed by blue star-forming objects there is also a statistical significant difference beteween the bimodal color distributions of the two samples of  galaxies.
We estimate a cluster total mass in the range $1.2 \ 10^{14}-3.5 \ 10^{14} \ M_{\odot}$ for bias factor $1<b<3$, adapting  a method already used for spectroscopic data at higher $z$ \cite{steidel}. 
This is the typical mass range for a poor cluster.

We have then checked the 1Ms exposure performed by Chandra satellite in the 0.4-3 keV interval to compute the X-ray emission. We find that the emission is divided in three separated clumps.
From the observed flux we obtain a total emission $L_X \sim0.5 \cdot 10^{43} erg \ s^{-1}$ (in the interval 2-10 keV), assuming a thermal spectrum with T=3 keV and abundance $Z=0.2 Z_{\odot}$. This total luminosity is lower than expected for a cluster of this mass and richness ($L_X > 10^{43} erg \ s^{-1}$). 
The low X-ray luminosity and its irregular morphology suggest that the group/poor cluster has not yet reached its virial equilibrium.

\begin{figure}
\includegraphics[width=8cm]{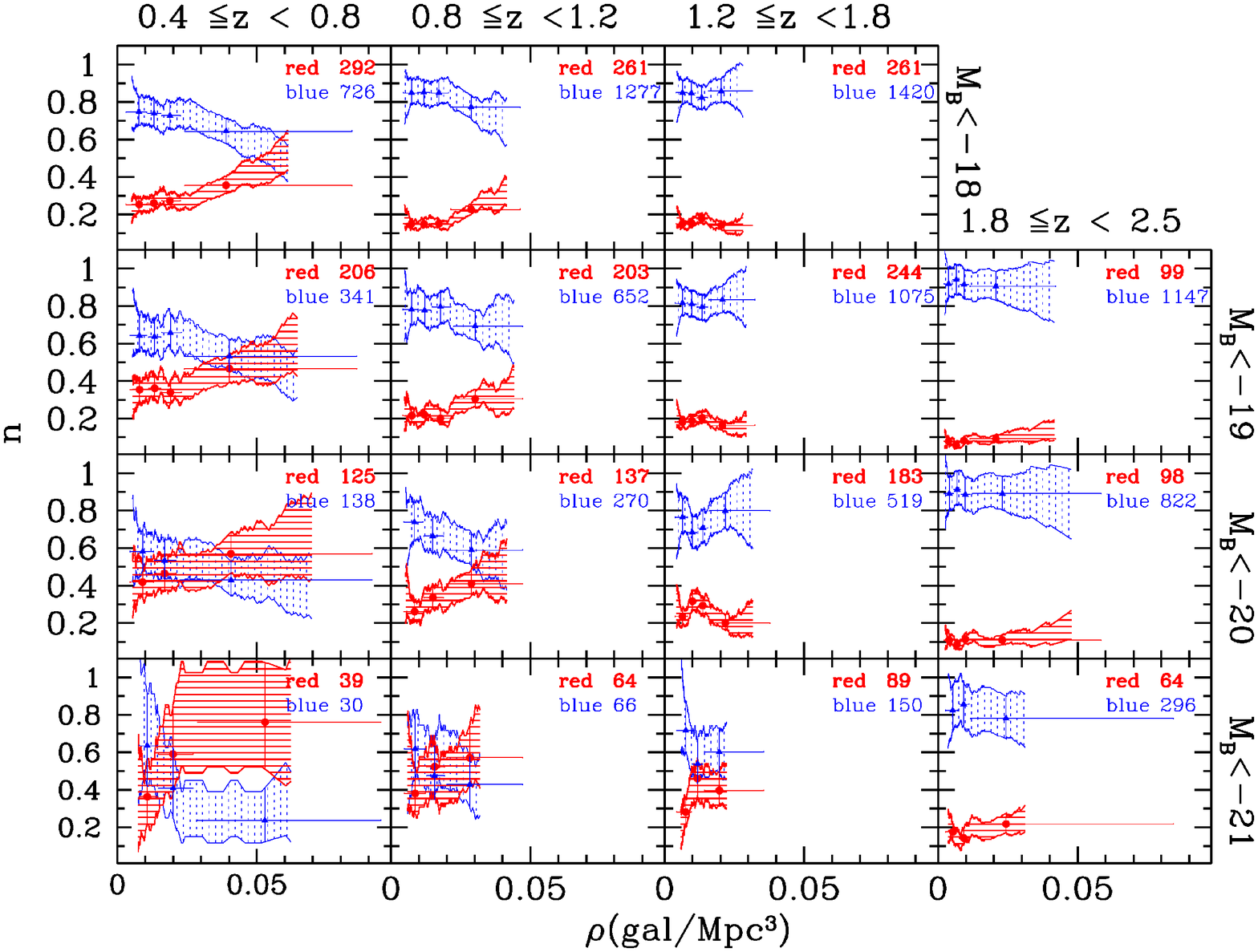}
\includegraphics[width=5cm]{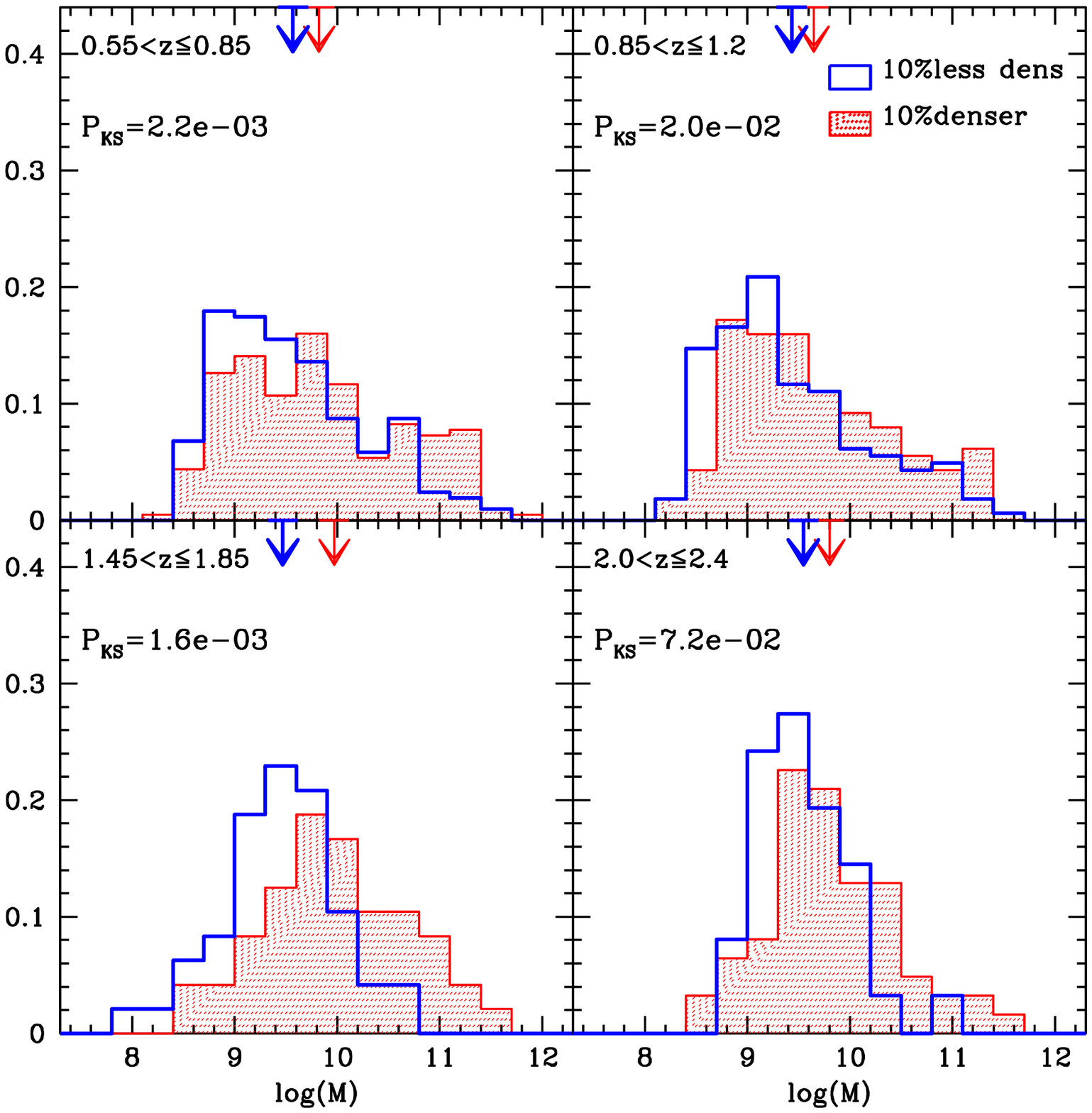}
\caption{Left: Fraction of blue and red galaxies with density at different redshifts and luminosities. Right: Distribution of total stellar mass for galaxies in the structures (shaded histogram) and in low density regions at various redshifts. In each panel it is shown the probability, according to a Kolmogorov-Smirnov test, that the two distributions are drawn from the same sample.}
\end{figure}

We can calculate the fraction of red and blue galaxies for different rest frame B  magnitudes in four redshift intervals. The red fraction increases at decreasing B magnitude, while, at fixed luminosity, it increases with decreasing redshift. At $z>1.5$ even the highest luminosity galaxies are blue, star-forming objects (Fig. 2), as it has been shown with a shallower sample of spectroscopic data in the VVDS ~\cite{cucciati06}. 
The total stellar masses have been estimated with the same procedure that yields the photometric redshift ($\chi^2$ minimization using the observed SED and template spectra, as described in ~\cite{fontana00}). In the four panels of Fig. 2 (right) we present the galaxy total stellar mass distribution in four redshift intervals.  At  every redshift, in agreement with a hierarchical clustering scenario, the galaxies at high environmental density have a distribution that peaks at higher masses with respect to 'field' galaxies.
\section{Conclusions}

We have estimated the galaxy volume density using photometric redshifts in the GOODS-MUSIC catalogue. We selected high-density peaks up to $z$=2.5. Among them we find a structure at redshift 1.6 with properties of a forming cluster of galaxies. A population of red galaxies is present in all of them. The slope of the colour-magnitude relation of this red population is comparable to the slope measured for lower $z$ structures. We have also  analyzed the variation with density of the distribution of galaxy colours, which is related with the distribution of galaxy types. We find that the red galaxy segregation at high densities is stronger at low redshift and at high luminosities but it gets weaker for increasing $z$. At the highest $z$ probed, a blue and star forming population prevails at every density and luminosity value. These results show that the trend previously obtained using a sample of spectroscopic data \cite{cucciati06} extends to higher redshifts. We have also found that massive galaxies prevails, at every redshift, in dense environments. With the help of photometric redshifts it will be possible to detect overdensities and observe the onset and growth of the bimodality of galaxy types distribution, at the highest reachable redshifts. This will allow to constrain models for galaxy evolution in clustered environments.

\bibliographystyle{varenna}

\end{document}